\documentclass{article}

\usepackage[final,nonatbib]{neurips_2023}

\usepackage[utf8]{inputenc} 
\usepackage[T1]{fontenc}    
\usepackage{hyperref}       
\usepackage{url}            
\usepackage{booktabs}       
\usepackage{amsfonts}       
\usepackage{nicefrac}       
\usepackage{microtype}      
\usepackage{xcolor}         
\usepackage{romannum} 
\usepackage{graphicx} 
\usepackage{biblatex}
\usepackage{subcaption}
\usepackage{amsmath}
\usepackage{amssymb}
\usepackage{bm}
\title{Semantic Map Guided Synthesis of Wireless Capsule Endoscopy Images using Diffusion Models}

\author{Haejin Lee \textsuperscript{\rm 1} \\
\And
Jeongwoo Ju \textsuperscript{\rm 1} \\
\And
Jonghyuck Lee \textsuperscript{\rm 1,2} \\
\textsuperscript{\rm 2} Seoreu Co., Ltd.\\
Busan 46288, Korea\\
\texttt{jhlee@seoreu.com}
\And
Yeoun Joo Lee \textsuperscript{\rm 1,3} \\
\textsuperscript{\rm 3} Department  of  Pediatrics\\
Pusan  National  University  School  of  Medicine\\
Pusan  National  University  Yangsan Hospital\\
Yangsan 50612, Korea\\
\texttt{moonmissing@gmail.com}
\And
Heechul Jung \textsuperscript{\rm 1,4} \\
\textsuperscript{\rm 4} Department of Artificial Intelligence\\
Kyungpook National University\\
Daegu 41566, Korea\\
\texttt{heechul@knu.ac.kr}\\
\And
\\
\textsuperscript{\rm 1} Captos Co., Ltd.\\
Yangsan 50602, Korea\\
\texttt{\{seareale,veryju,jhlee,yeounjoolee,heechul\}@captos.co.kr}\\
}

\begin{document}

\maketitle

\begin{abstract}
Wireless capsule endoscopy (WCE) is a non-invasive method for visualizing the gastrointestinal (GI) tract, crucial for diagnosing GI tract diseases. However, interpreting WCE results can be time-consuming and tiring. Existing studies have employed deep neural networks (DNNs) for automatic GI tract lesion detection, but acquiring sufficient training examples, particularly due to privacy concerns, remains a challenge. Public WCE databases lack diversity and quantity. To address this, we propose a novel approach leveraging generative models, specifically the diffusion model (DM), for generating diverse WCE images. Our model incorporates semantic map resulted from visualization scale (VS) engine, enhancing the controllability and diversity of generated images. We evaluate our approach using visual inspection and visual Turing tests, demonstrating its effectiveness in generating realistic and diverse WCE images.
\end{abstract}

\section{Introduction}
WCE \cite{iddan2000wireless} is a non-invasive method for viewing the GI tract, aiding in GI tract disease diagnosis. However, reading WCE results can be time-consuming and lead to fatigue. Studies \cite{soffer2020deep,barash2021ulcer,yang2020future,aoki2021automatic} have worked on automatic GI tract lesion detection using deep neural networks (DNNs), which require extensive training examples like ImageNet \cite{deng2009imagenet}. However, privacy concerns make gathering enough examples challenging \cite{voigt2017eu}. Morevover, public WCE databases \cite{koulaouzidis2017kid,Smedsrud2021} lack diversity and sufficient examples. Generating data using generative models \cite{diamantis2023intestine} may be a solution.
Generative models, especially deep learning-based, excel in generating high-quality data in text \cite{iqbal2022survey}, image \cite{elasri2022image}, and audio \cite{huzaifah2021deep,oord2016wavenet}. They address issues like class imbalance \cite{sampath2021survey} and example augmentation \cite{antoniou2017data,frid2018gan}, relevant in the medical domain.
Most works, including medical image datasets, rely on generative adversarial networks (GANs) \cite{goodfellow2020generative}. Denoising diffusion probabilistic models (DDPMs) \cite{ho2020denoising} capture real image variations effectively, potentially outperforming GANs \cite{dhariwal2021diffusion}. Recent image synthesis efforts favor DDPM approaches \cite{muller2022diffusion,kazerouni2022diffusion,yang2022diffusion}.
In the medical domain, attention to WCE is limited compared to other modalities. Diamantis et al. \cite{diamantis2022endovae,diamantis2023intestine,diamantis2022endovae,diamantis2019towards} and Valts et al. \cite{vats2023evaluating,vats2023changes} addressed WCE image synthesis, but more sophisticated techniques for diversity and controllability are needed.
This study aims to generate diverse WCE images using a diffusion model with semantic segmentation map. 
We achieved controllability over the semantic map, which was readily obtained through the VS engine \cite{ju2022semantic}, enabling the generation of WCE images with desired semantic map.
Additionally, as an WCE image is composed of semantic area (e.g. clean area, dark area, the other area where floating debris or bubbles occupy) such that one can easily notice those area at a glance. Also, considering the fact that lesion can only appear inside clean area and the rest area can exhibit different appearance in a various way, controlling semantic area is a key technique to achieve greater diversity.

Evaluation included visual inspection and visual turing test.
Contributions:\
\romannum{1}) Pioneering use of DM for synthesized WCE images.\
\romannum{2}) Incorporation of floats/bubbles and darkness area as controllable factors using VS results.

\section{Methods}
\begin{figure*}
\includegraphics[width=\textwidth]{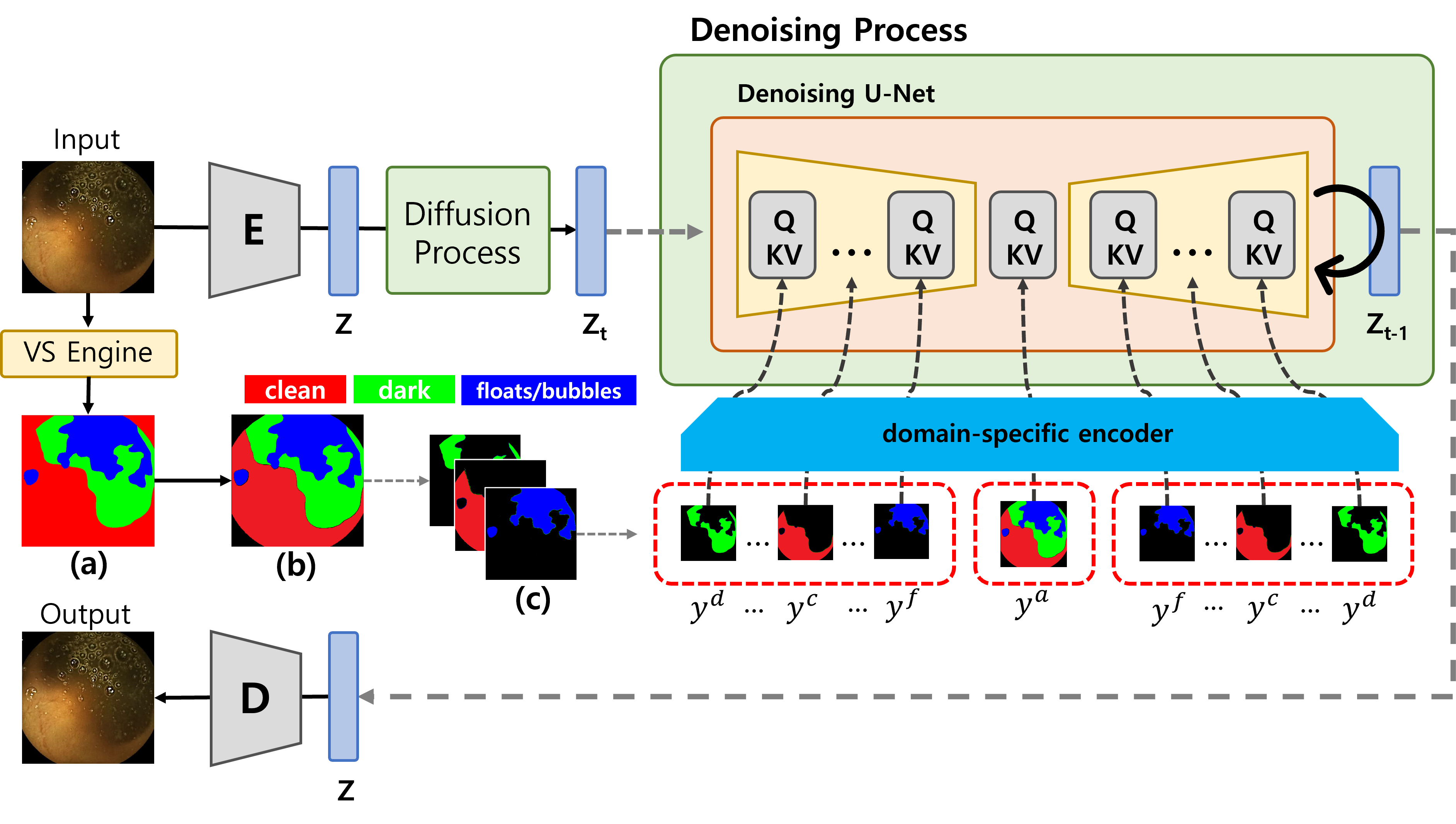}
\caption{Flowchart for training the Diffusion Model for WCE image generation. We start by obtaining the segmentation map from the VS engine. This map is then divided into class-specific maps. These individual maps are subsequently integrated into the Encoder, Decoder, and middle layer of the Denoising U-Net.} \label{fig:flowchart}
\vspace{-3mm}
\end{figure*}

\subsection{Kvasir-capsule Dataset}
In this study, we used the Kvasir-Capsule dataset, which is the largest publicly available collection of WCE images captured by PillCam (Medtronic, Minneapolis, MN, USA). The dataset consists of 47,238 labeled images and 43 labeled videos, along with 74 unlabeled videos. It includes images corresponding to multiple organs, such as the stomach, small bowel (SB) and colon. We used all labeled images to train our latent diffusion model (LDM).


\subsection{Segmentation Mask}
Ju et al. \cite{ju2022semantic} developed visualization scale (VS) engine designed for semantic segmentation to assess the clean mucosa area in small bowel images captured by WCE. The target classes for segmentation are darkness (absence of light), floats/bubbles (bubbles and floating debris), and clean (clear visual area). The overall performance, measured by the Dice index, was 0.9457. The comparison between physicians and the AI engine demonstrated good agreement in accessing the clean area \cite{ju2023clean}.

In this study, we fed the images to the VS engine and obtained a semantic segmentation map. However, the VS engine mistakenly regarded the blank areas at the four corners of a WCE image as clean areas (Figure \ref{fig:flowchart}.(a)). To address this issue, we manually filtered out the corners from the semantic map and reassigned them as an extra class (Figure \ref{fig:flowchart}.(b)). Additionally, we split the single-channel semantic map into three channels, where each channel depicts a single semantic class (Figure \ref{fig:flowchart}.(c)). As a result, we obtained three channels based on the single semantic map.

\subsection{Mask-conditioned Latent Diffusion Models}
Rombach et al. \cite{rombach2022high} introduced LDM which is computationally faster version of vanilla DM while keeping synthesis performance. Moreover, one of key technique for generative model is modelling conditional distribution, $p(\cdot|\bm{y})$, they proposed conditioning mechanism with following loss function,
$L_{DM}=\mathbb{E}_{\bm{z},\bm{y},\bm{\epsilon} \sim \mathcal{N}(\bm{0},\bm{I}),t} \left[\big\| \bm{\epsilon} - f_{\bm{\theta}}(\bm{z}_{t},t,\tau_{\bm{\theta}}(\bm{y})) \big\|_{2}^{2}\right]$,
where $t$ denotes time uniformly sampled from $\{1,...,T\}$, $f_{\bm{\theta}}(*,t,\tau_{\bm{\theta}}(\bm{y}))$ is conditional denoising autoencoder,
$\tau_{\bm{\theta}}(\bm{y})$ is a domain-specific encoder, $\bm{z}_{t}$ is the noisy variant of latent variable $\bm{z}$ at time $t$, and $\bm{y}$ is conditional variable. 
$f_{\theta}$ is basically implemented by introducing the backbone with U-Net \cite{ronneberger2015u} and cross-attention mechanism \cite{vaswani2017attention}. Moreover, feeding conditional variable via $\tau_{\bm{\theta}}$ into the intermediate layers of UNet is the core of modelling $p(\bm{z}|\bm{y})$.
In contrast to existing methods where $\bm{y}$ is fixed as a single segmentation map, we divided $\bm{y}$ into multiple semantic maps. Specifically, $\bm{y}$ is composed of maps sampled from $\{\bm{y}^{d},\bm{y}^{c},\bm{y}^{f},\bm{y}^{a}\}$, where $\bm{y}^{d},\bm{y}^{c},\bm{y}^{f},\bm{y}^{a}$ denote the segmentation maps for dark, clean, floats/bubbles, and all areas respectively. When implementing the denoising U-Net, we sequentially fed samples from $\{\bm{y}^{d},\bm{y}^{c},\bm{y}^{f}\}$ into the U-Net encoder in a specified order, $\bm{y}^{a}$ into the middle layer, and the reversed samples (which is the exact reverse order of semantic maps for the encoder) into the decoder. By doing so, we expected U-Net to learn sequentially on how to generate each semantic area.

\section{Results}
\begin{figure*}[htp]
\begin{center}
\begin{tabular}{ccccc}
\begin{minipage}{27mm}\centering{\textbf{mask1}}\end{minipage}
\begin{minipage}{27mm}\centering{\textbf{mask2}}\end{minipage}
\begin{minipage}{27mm}\centering{\textbf{mask3}}\end{minipage}
\begin{minipage}{27mm}\centering{\textbf{mask4}}\end{minipage}
\begin{minipage}{27mm}\centering{\textbf{mask5}}\end{minipage}
\vspace{2mm}
\\
\begin{minipage}{27mm}\includegraphics[width=27mm]{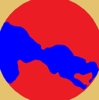}\end{minipage}
\begin{minipage}{27mm}\includegraphics[width=27mm]{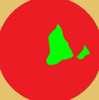}\end{minipage}
\begin{minipage}{27mm}\includegraphics[width=27mm]{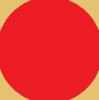}\end{minipage}
\begin{minipage}{27mm}\includegraphics[width=27mm]{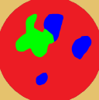}\end{minipage}
\begin{minipage}{27mm}\includegraphics[width=27mm]{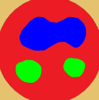}\end{minipage}
\vspace{2mm}
\\
\begin{minipage}{27mm}\includegraphics[width=27mm]{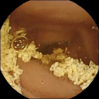}\end{minipage}
\begin{minipage}{27mm}\includegraphics[width=27mm]{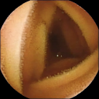}\end{minipage}
\begin{minipage}{27mm}\includegraphics[width=27mm]{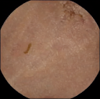}\end{minipage}
\begin{minipage}{27mm}\includegraphics[width=27mm]{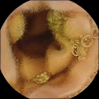}\end{minipage}
\begin{minipage}{27mm}\includegraphics[width=27mm]{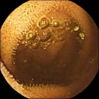}\end{minipage}
\vspace{2mm}
\\
\begin{minipage}{27mm}\includegraphics[width=27mm]{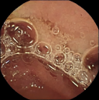}\end{minipage}
\begin{minipage}{27mm}\includegraphics[width=27mm]{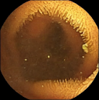}\end{minipage}
\begin{minipage}{27mm}\includegraphics[width=27mm]{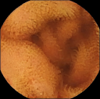}\end{minipage}
\begin{minipage}{27mm}\includegraphics[width=27mm]{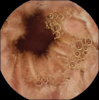}\end{minipage}
\begin{minipage}{27mm}\includegraphics[width=27mm]{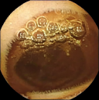}\end{minipage}
\vspace{2mm}
\\
\begin{minipage}{27mm}\includegraphics[width=27mm]{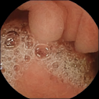}\end{minipage}
\begin{minipage}{27mm}\includegraphics[width=27mm]{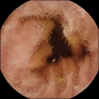}\end{minipage}
\begin{minipage}{27mm}\includegraphics[width=27mm]{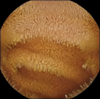}\end{minipage}
\begin{minipage}{27mm}\includegraphics[width=27mm]{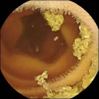}\end{minipage}
\begin{minipage}{27mm}\includegraphics[width=27mm]{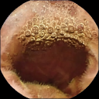}\end{minipage}
\vspace{2mm}
\\
\begin{minipage}{27mm}\includegraphics[width=27mm]{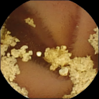}\end{minipage}
\begin{minipage}{27mm}\includegraphics[width=27mm]{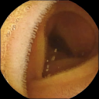}\end{minipage}
\begin{minipage}{27mm}\includegraphics[width=27mm]{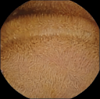}\end{minipage}
\begin{minipage}{27mm}\includegraphics[width=27mm]{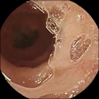}\end{minipage}
\begin{minipage}{27mm}\includegraphics[width=27mm]{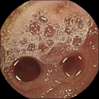}\end{minipage}
\vspace{2mm}
\\
\begin{minipage}{27mm}\includegraphics[width=27mm]{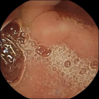}\end{minipage}
\begin{minipage}{27mm}\includegraphics[width=27mm]{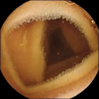}\end{minipage}
\begin{minipage}{27mm}\includegraphics[width=27mm]{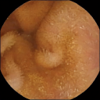}\end{minipage}
\begin{minipage}{27mm}\includegraphics[width=27mm]{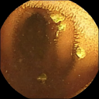}\end{minipage}
\begin{minipage}{27mm}\includegraphics[width=27mm]{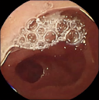}\end{minipage}
\vspace{2mm}
\\
\end{tabular}
\caption{The segmentation maps used in WCE image generation (top row). The masks 1 to 3 were obtained from real images, while masks 4 to 5 were created by us. The generated images corresponding to each map are shown in the subsequent rows. \textbf{Beige-brown: blank area,  red: clean area,  green: dark area,  blue: floats/bubbles area.}}
\label{fig:generation}
\end{center}
\end{figure*}

\paragraph{VTT} Five gastroenterologists with over 5 years of experience in capsule endoscopy interpretation conducted a visual Turing test using a total of 160 images, including 80 real and 80 fake images. The results showed an average accuracy of 0.64 in correctly identifying real images as real, and a 0.662 accuracy in incorrectly identifying fake images as real. This indicates that the quality of the generated WCE is perceived to be comparable to real images, as evidenced by the similar rates at which experts classified real and fake images as real during the assessment of image quality.
\paragraph{Visual Inspection} Figure \ref{fig:generation} displays images generated with varying random seeds when an arbitrary segmentation mask is provided (top row). It can be observed that different areas are created in accordance with the regions outlined in the segmentation mask, including dark, clean, and floats/bubbles regions. Notably, despite the fact that mask 5 represent semantic maps that are challenging to exist in reality (we created), our DM successfully generated corresponding areas in accordance with these maps.

\section{Related Work}
\paragraph{Generating WCE images} Diamantis et al. \cite{diamantis2019towards} attempted WCE image generation with GANs. However, their study lacked extensive experiments, only offering binary classification results. Their subsequent work, EndoVAE \cite{diamantis2022endovae}, used a simple architecture but still lacked thorough experiments, focusing only on binary classification. Their latest work, TIDE \cite{diamantis2023intestine}, improved VAE by incorporating multiscale blocks (MSB) for high-quality image synthesis, outperforming GAN variants.
Vats et al. \cite{vats2023evaluating} aimed to create a WCE atlas and simulate disease progression for educational purposes. They built an embedding space using StyleGAN2 due to limited annotated WCE data. Although their synthetic images were high-quality, attribute discovery involved human validation. However, our work has demonstrated the ability to control semantic areas, generating high-quality WCE images without the need for human intervention.
\paragraph{Diffusion-based Models} Sohl-Dickstein et al. \cite{sohl2015deep} introduced DM inspired by statistical physics. Ho et al. \cite{ho2020denoising} demonstrated DM's capability for synthetic image generation. However, DM's high computational cost led to the development of LDM by Rombach et al. \cite{rombach2022high}, achieving similar quality with less resource consumption. LDM's innovation lies in utilizing DM in the latent space of a pretrained autoencoder and incorporating cross-attention layers for conditional image generation with text input. We chose LDM as our base WCE image generator.

\section{Conclusion}
In this study, we introduced a novel approach for generating diverse WCE images using a DM with semantic segmentation maps. By obtaining controllability over the semantic map, we demonstrated the ability to generate WCE images with desired appearances, eliminating the need for additional annotations. The evaluation, including visual inspection and a visual Turing test, confirmed the high quality and realism of the generated images. Additionally, we incorporated floats/bubbles and darkness area as controllable factors, enhancing the model's versatility. This work paves the way for improved WCE image synthesis, providing a valuable resource for medical research and education.

\newpage
\printbibliography

@article{iddan2000wireless,
  title={Wireless capsule endoscopy},
  author={Iddan, Gavriel and Meron, Gavriel and Glukhovsky, Arkady and Swain, Paul},
  journal={Nature},
  volume={405},
  number={6785},
  pages={417--417},
  year={2000},
  publisher={Nature Publishing Group UK London}
}

@article{barash2021ulcer,
  title={Ulcer severity grading in video capsule images of patients with Crohn’s disease: An ordinal neural network solution},
  author={Barash, Yiftach and Azaria, Liran and Soffer, Shelly and Yehuda, Reuma Margalit and Shlomi, Oranit and Ben-Horin, Shomron and Eliakim, Rami and Klang, Eyal and Kopylov, Uri},
  journal={Gastrointestinal Endoscopy},
  volume={93},
  number={1},
  pages={187--192},
  year={2021},
  publisher={Elsevier}
}

@article{yang2020future,
  title={The future of capsule endoscopy: The role of artificial intelligence and other technical advancements},
  author={Yang, Young Joo},
  journal={Clinical Endoscopy},
  volume={53},
  number={4},
  pages={387--394},
  year={2020},
  publisher={Korean Society of Gastrointestinal Endoscopy}
}

@article{aoki2021automatic,
  title={Automatic detection of various abnormalities in capsule endoscopy videos by a deep learning-based system: a multicenter study},
  author={Aoki, Tomonori and Yamada, Atsuo and Kato, Yusuke and Saito, Hiroaki and Tsuboi, Akiyoshi and Nakada, Ayako and Niikura, Ryota and Fujishiro, Mitsuhiro and Oka, Shiro and Ishihara, Soichiro and others},
  journal={Gastrointestinal Endoscopy},
  volume={93},
  number={1},
  pages={165--173},
  year={2021},
  publisher={Elsevier}
}

@article{soffer2020deep,
  title={Deep learning for wireless capsule endoscopy: a systematic review and meta-analysis},
  author={Soffer, Shelly and Klang, Eyal and Shimon, Orit and Nachmias, Noy and Eliakim, Rami and Ben-Horin, Shomron and Kopylov, Uri and Barash, Yiftach},
  journal={Gastrointestinal endoscopy},
  volume={92},
  number={4},
  pages={831--839},
  year={2020},
  publisher={Elsevier}
}

@article{voigt2017eu,
  title={The eu general data protection regulation (gdpr)},
  author={Voigt, Paul and Von dem Bussche, Axel},
  journal={A Practical Guide, 1st Ed., Cham: Springer International Publishing},
  volume={10},
  number={3152676},
  pages={10--5555},
  year={2017},
  publisher={Springer}
}

@inproceedings{deng2009imagenet,
  title={Imagenet: A large-scale hierarchical image database},
  author={Deng, Jia and Dong, Wei and Socher, Richard and Li, Li-Jia and Li, Kai and Fei-Fei, Li},
  booktitle={2009 IEEE conference on computer vision and pattern recognition},
  pages={248--255},
  year={2009},
  organization={Ieee}
}

@article{koulaouzidis2017kid,
  title={KID Project: an internet-based digital video atlas of capsule endoscopy for research purposes},
  author={Koulaouzidis, Anastasios and Iakovidis, Dimitris K and Yung, Diana E and Rondonotti, Emanuele and Kopylov, Uri and Plevris, John N and Toth, Ervin and Eliakim, Abraham and Johansson, Gabrielle Wurm and Marlicz, Wojciech and others},
  journal={Endoscopy international open},
  volume={5},
  number={06},
  pages={E477--E483},
  year={2017},
  publisher={{\copyright} Georg Thieme Verlag KG}
}

@article{Smedsrud2021,
  title = {{Kvasir-Capsule, a video capsule endoscopy dataset}},
  author = {
    Smedsrud, Pia H and Thambawita, Vajira and Hicks, Steven A and
    Gjestang, Henrik and Nedrejord, Oda Olsen and N{\ae}ss, Espen and
    Borgli, Hanna and Jha, Debesh and Berstad, Tor Jan Derek and
    Eskeland, Sigrun L and Lux, Mathias and Espeland, H{\aa}vard and
    Petlund, Andreas and Nguyen, Duc Tien Dang and Garcia-Ceja, Enrique and
    Johansen, Dag and Schmidt, Peter T and Toth, Ervin and
    Hammer, Hugo L and de Lange, Thomas and Riegler, Michael A and
    Halvorsen, P{\aa}l
  },
  doi = {10.1038/s41597-021-00920-z},
  journal = {Scientific Data},
  number = {1},
  pages = {142},
  volume = {8},
  year = {2021}
}

@article{vaswani2017attention,
  title={Attention is all you need},
  author={Vaswani, Ashish and Shazeer, Noam and Parmar, Niki and Uszkoreit, Jakob and Jones, Llion and Gomez, Aidan N and Kaiser, {\L}ukasz and Polosukhin, Illia},
  journal={Advances in neural information processing systems},
  volume={30},
  year={2017}
}

@misc{ho2020denoising,
      title={Denoising Diffusion Probabilistic Models}, 
      author={Jonathan Ho and Ajay Jain and Pieter Abbeel},
      year={2020},
      eprint={2006.11239},
      archivePrefix={arXiv},
      primaryClass={cs.LG}
}

@article{dhariwal2021diffusion,
  title={Diffusion models beat gans on image synthesis},
  author={Dhariwal, Prafulla and Nichol, Alexander},
  journal={Advances in Neural Information Processing Systems},
  volume={34},
  pages={8780--8794},
  year={2021}
}

@inproceedings{sohl2015deep,
  title={Deep unsupervised learning using nonequilibrium thermodynamics},
  author={Sohl-Dickstein, Jascha and Weiss, Eric and Maheswaranathan, Niru and Ganguli, Surya},
  booktitle={International Conference on Machine Learning},
  pages={2256--2265},
  year={2015},
  organization={PMLR}
}

@inproceedings{rombach2022high,
  title={High-resolution image synthesis with latent diffusion models},
  author={Rombach, Robin and Blattmann, Andreas and Lorenz, Dominik and Esser, Patrick and Ommer, Bj{\"o}rn},
  booktitle={Proceedings of the IEEE/CVF Conference on Computer Vision and Pattern Recognition},
  pages={10684--10695},
  year={2022}
}

@article{kazerouni2022diffusion,
  title={Diffusion models for medical image analysis: A comprehensive survey},
  author={Kazerouni, Amirhossein and Aghdam, Ehsan Khodapanah and Heidari, Moein and Azad, Reza and Fayyaz, Mohsen and Hacihaliloglu, Ilker and Merhof, Dorit},
  journal={arXiv preprint arXiv:2211.07804},
  year={2022}
}

@article{muller2022diffusion,
  title={Diffusion Probabilistic Models beat GANs on Medical Images},
  author={M{\"u}ller-Franzes, Gustav and Niehues, Jan Moritz and Khader, Firas and Arasteh, Soroosh Tayebi and Haarburger, Christoph and Kuhl, Christiane and Wang, Tianci and Han, Tianyu and Nebelung, Sven and Kather, Jakob Nikolas and others},
  journal={arXiv preprint arXiv:2212.07501},
  year={2022}
}

@article{yang2022diffusion,
  title={Diffusion models: A comprehensive survey of methods and applications},
  author={Yang, Ling and Zhang, Zhilong and Song, Yang and Hong, Shenda and Xu, Runsheng and Zhao, Yue and Shao, Yingxia and Zhang, Wentao and Cui, Bin and Yang, Ming-Hsuan},
  journal={arXiv preprint arXiv:2209.00796},
  year={2022}
}

@inproceedings{ronneberger2015u,
  title={U-net: Convolutional networks for biomedical image segmentation},
  author={Ronneberger, Olaf and Fischer, Philipp and Brox, Thomas},
  booktitle={Medical Image Computing and Computer-Assisted Intervention--MICCAI 2015: 18th International Conference, Munich, Germany, October 5-9, 2015, Proceedings, Part III 18},
  pages={234--241},
  year={2015},
  organization={Springer}
}

@article{oord2016wavenet,
  title={Wavenet: A generative model for raw audio},
  author={Oord, Aaron van den and Dieleman, Sander and Zen, Heiga and Simonyan, Karen and Vinyals, Oriol and Graves, Alex and Kalchbrenner, Nal and Senior, Andrew and Kavukcuoglu, Koray},
  journal={arXiv preprint arXiv:1609.03499},
  year={2016}
}

@article{antoniou2017data,
  title={Data augmentation generative adversarial networks},
  author={Antoniou, Antreas and Storkey, Amos and Edwards, Harrison},
  journal={arXiv preprint arXiv:1711.04340},
  year={2017}
}

@article{frid2018gan,
  title={GAN-based synthetic medical image augmentation for increased CNN performance in liver lesion classification},
  author={Frid-Adar, Maayan and Diamant, Idit and Klang, Eyal and Amitai, Michal and Goldberger, Jacob and Greenspan, Hayit},
  journal={Neurocomputing},
  volume={321},
  pages={321--331},
  year={2018},
  publisher={Elsevier}
}

@article{huzaifah2021deep,
  title={Deep generative models for musical audio synthesis},
  author={Huzaifah, Muhammad and Wyse, Lonce},
  journal={Handbook of Artificial Intelligence for Music: Foundations, Advanced Approaches, and Developments for Creativity},
  pages={639--678},
  year={2021},
  publisher={Springer}
}

@article{sampath2021survey,
  title={A survey on generative adversarial networks for imbalance problems in computer vision tasks},
  author={Sampath, Vignesh and Maurtua, I{\~n}aki and Aguilar Martin, Juan Jose and Gutierrez, Aitor},
  journal={Journal of big Data},
  volume={8},
  pages={1--59},
  year={2021},
  publisher={Springer}
}

@article{elasri2022image,
  title={Image Generation: A Review},
  author={Elasri, Mohamed and Elharrouss, Omar and Al-Maadeed, Somaya and Tairi, Hamid},
  journal={Neural Processing Letters},
  volume={54},
  number={5},
  pages={4609--4646},
  year={2022},
  publisher={Springer}
}

@article{goodfellow2020generative,
  title={Generative adversarial networks},
  author={Goodfellow, Ian and Pouget-Abadie, Jean and Mirza, Mehdi and Xu, Bing and Warde-Farley, David and Ozair, Sherjil and Courville, Aaron and Bengio, Yoshua},
  journal={Communications of the ACM},
  volume={63},
  number={11},
  pages={139--144},
  year={2020},
  publisher={ACM New York, NY, USA}
}

@article{iqbal2022survey,
  title={The survey: Text generation models in deep learning},
  author={Iqbal, Touseef and Qureshi, Shaima},
  journal={Journal of King Saud University-Computer and Information Sciences},
  volume={34},
  number={6},
  pages={2515--2528},
  year={2022},
  publisher={Elsevier}
}

@article{ju2022semantic,
  title={Semantic Segmentation Dataset for AI-Based Quantification of Clean Mucosa in Capsule Endoscopy},
  author={Ju, Jeong-Woo and Jung, Heechul and Lee, Yeoun Joo and Mun, Sang-Wook and Lee, Jong-Hyuck},
  journal={Medicina},
  volume={58},
  number={3},
  pages={397},
  year={2022},
  publisher={MDPI}
}

@article{ju2023clean,
  title={Clean mucosal area detection of gastroenterologists versus artificial intelligence in small bowel capsule endoscopy},
  author={Ju, Jeongwoo and Oh, Hyun Sook and Lee, Yeoun Joo and Jung, Heechul and Lee, Jong-Hyuck and Kang, Ben and Choi, Sujin and Kim, Ji Hyun and Kim, Kyeong Ok and Chung, Yun Jin},
  journal={Medicine},
  volume={102},
  number={6},
  year={2023},
  publisher={Wolters Kluwer Health}
}

@inproceedings{diamantis2019towards,
  title={Towards the substitution of real with artificially generated endoscopic images for CNN training},
  author={Diamantis, Dimitris E and Zacharia, Athena E and Iakovidis, Dimitris K and Koulaouzidis, Anastasios},
  booktitle={2019 IEEE 19th International Conference on Bioinformatics and Bioengineering (BIBE)},
  pages={519--524},
  year={2019},
  organization={IEEE}
}

@inproceedings{diamantis2022endovae,
  title={EndoVAE: Generating Endoscopic Images with a Variational Autoencoder},
  author={Diamantis, Dimitrios E and Gatoula, Panagiota and Iakovidis, Dimitris K},
  booktitle={2022 IEEE 14th Image, Video, and Multidimensional Signal Processing Workshop (IVMSP)},
  pages={1--5},
  year={2022},
  organization={IEEE}
}

@article{diamantis2023intestine,
  title={This Intestine Does Not Exist: Multiscale Residual Variational Autoencoder for Realistic Wireless Capsule Endoscopy Image Generation},
  author={Diamantis, Dimitrios E and Gatoula, Panagiota and Koulaouzidis, Anastasios and Iakovidis, Dimitris K},
  journal={arXiv preprint arXiv:2302.02150},
  year={2023}
}

@inproceedings{vats2023changes,
  title={This changes to that: Combining causal and non-causal explanations to generate disease progression in capsule endoscopy},
  author={Vats, Anuja and Mohammed, Ahmed and Pedersen, Marius and Wiratunga, Nirmalie},
  booktitle={ICASSP 2023-2023 IEEE International Conference on Acoustics, Speech and Signal Processing (ICASSP)},
  pages={1--5},
  year={2023},
  organization={IEEE}
}

@article{vats2023evaluating,
  title={Evaluating clinical diversity and plausibility of synthetic capsule endoscopic images},
  author={Vats, Anuja and Pedersen, Marius and Mohammed, Ahmed and Hovde, {\O}istein},
  journal={arXiv preprint arXiv:2301.06366},
  year={2023}
}
\end{document}